\documentclass[twocolumn,twocolappendix]{aastex63}
\pdfoutput=1
\newcommand{\sgra}{Sgr~A$^*$}

\usepackage{graphicx} 
\usepackage{amsmath}
\usepackage{subfigure}
\usepackage{amssymb}

\begin{document}

\title{THE UNEXPECTED LACK OF ASYMMETRY IN THE HORIZON-SCALE IMAGE OF SAGITTARIUS A*}

\author[0000-0003-4440-8924]{J.Cole Faggert}
\affiliation{School of Physics, Georgia Institute of Technology, 837 State St NW, Atlanta, GA 30332, USA}

\author[0000-0003-4413-1523]{Feryal \"Ozel}
\affiliation{School of Physics, Georgia Institute of Technology, 837 State St NW, Atlanta, GA 30332, USA}

\author[0000-0003-1035-3240]{Dimitrios Psaltis}
\affiliation{School of Physics, Georgia Institute of Technology, 837 State St NW, Atlanta, GA 30332, USA}

\begin{abstract}
The ring-like images of the two supermassive black holes captured by the Event Horizon Telescope (EHT) provide powerful probes of the physics of accretion flows at horizon scales. Specifically, the brightness asymmetry in the images carries information about the angular velocity profile of the inner accretion flow and the inclination of the observer, owing to the Doppler boosts photons experience at their site of emission. 
In this paper, we develop a method for quantifying the brightness asymmetry of black-hole images in the Fourier domain, which can be measured directly from interferometric data. We apply this method to current EHT data and find that the image of Sagittarius~A* (\sgra) has an unusually low degree of asymmetry that is even lower than that inferred for M87. We then use a covariant semi-analytic model to obtain constraints on the inclinations and velocity profiles of the inner accretion flow for \sgra. We find that the lack of significant brightness asymmetry forces the observer inclination to uncomfortably small values ($6-10^\circ$), if the plasma velocity follows Keplerian profiles. Alternatively, larger inclination angles can be accommodated if the plasma velocities are significantly sub-Keplerian and the black hole is not spinning rapidly. 
\end{abstract}


\section{Introduction} \label{sec:intro}

Images of black holes obtained with the Event Horizon Telescope (EHT) have provided an exciting new avenue for exploring the physics of black holes and the plasmas of their accretion flow. The two closest EHT targets, the supermassive black holes (SMBHs) M87  and Sagittarius A$^*$ (\sgra), are fed by a class of accretion flows known as Radiatively Inefficient Accretion Flows. They are characterized by their low luminosity and high temperature, which lead to large scale-height disks. These sources become optically thin near the event horizon scales at millimeter wavelengths \citep{OzelOpticallyThin}, a fact that enables the EHT to reveal their ring-like images~\citep{M87PaperI,SgrAPaperI}.

The characteristics of these images offer the opportunity to learn about the physics at play in black hole accretion disks. In particular, the large-scale asymmetry in the ring-like feature probes plasma velocities of the inner accretion flow, owing to a Doppler signature that affects the characteristics of the image. Doppler boosting causes the radiative intensity observed from the plasma orbiting the black hole that moves toward the observer to be higher  relative to the intensity from the plasma that is moving away from the observer. This introduces a substantial brightness asymmetry around the ring-like image. With orbital velocities that become a significant fraction of the speed of light within several gravitational radii of the black hole, Doppler boosting is expected to have a significant impact on the images. The degree of these effects  depends on both the velocity profile of the accretion flow as well as the inclination of the observer with respect to the black hole spin axis.

It is important to note that the brightness asymmetry persists despite the strong gravitational lensing experienced by photons that contribute to the EHT image (at 230\;GHz). Indeed, 5-10\% of the flux corresponds to photons the geodesics of which crossed the equatorial plane of the accretion flow three times or more~\citep{Johannsen2012,BlackHoleImagesAsTestPlasma}, but this is not sufficient to reduce significantly the expected asymmetry due to Doppler effects. 

When analyzing the brightness asymmetry of the EHT targets, the black hole in M87 has the advantage of prior constraints on its inclination based on measurements of relativistic Doppler effects in its jet emission~\citep{Walker_Jets}. Assuming that the angular momentum of its accretion flow and its jet orientation are aligned, these measurements constrained the inclination to $\sim 17^{\circ}$ and suggested that there would be a low degree of asymmetry in the image, as the source would be viewed fairly face-on. Indeed, qualitatively, the reconstructed images by the EHT collaboration showed a ring-like image and not an asymmetric crescent that might have been expected if the source were viewed edge on. This was further explored quantitatively by \citet{Medeiros_2022}, who measured the brightness asymmetry in the EHT image of the M87 black hole and showed it to be relatively low, consistent with the inferred inclination.

Expectations for the inclination of Sgr~A$^*$ prior to the most recent EHT results were based upon several different arguments. From a formation perspective, the alignment (or lack thereof) of the galaxy's angular momentum and the spin of the black hole is not well understood. Models suggest that supermassive black holes that grow primarily from the accretion of the gas of the galaxy are more aligned with the angular momentum of the host galaxy, whereas those formed through mergers have more random orientations between the black hole spin and their host galaxy. 

From an observational standpoint, studies of the orbits and associated angular momentum vectors of the closest stars around Sgr~A$^*$ revealed the presence of nearly co-aligned orbits with an average inclination of $i=54 \pm 3.2$ \citep{BartkoSgrAStellar}. Initially, this inclination also appeared to be consistent with that inferred for Sgr~A$^*$ using the observed image size from the early pre-EHT observations~(\citealt{Psaltis2015}, see also \citealt{DexterSgr,HuangSgr,ShcherbakovSgr}. In contrast, astrometric observations with \texttt{GRAVITY} of the motion of the compact emitting regions around \sgra\ during infrared flares suggest a smaller inclination of $\sim 20-40^\circ$~\citep{Gravity2018,Gravity2020,Ball2021}.

Qualitative analyses of the azimuthal brightness profiles around the ring-like images of \sgra\ obtained by the EHT show a degree of brightness asymmetry that is much smaller than would be expected for relatively high inclinations \citep{SgrPaperIII}. In particular, the comparison of images to GRMHD models strongly disfavored models with inclination $>50^{\circ}$~\citep{SgrPaperV}. However, there are two complications that impact such comparisons. First, the presence of bright spots on the image, originating from the sparse image reconstruction techniques used, hampers the brightness asymmetry measure. Because the EHT obtains measurements in the Fourier domain, which are then converted to images through a variety of image reconstruction methods, image artifacts and assumptions used to construct the images can affect the quantitative measures of brightness asymmetry carried out in the image domain. Second, both the plasma velocities in the inner accretion flow and the observer inclination affect the degree of the asymmetry of a black hole image. As a result, constraints on the inclination of the source  can only be obtained and interpreted in combination with assumptions regarding the plasma velocities. 

To overcome the uncertainties potentially introduced by measurements of brightness asymmetry in the image domain, we develop in this paper a method of quantifying asymmetry directly in the Fourier domain, utilizing the complex visibilities measured by the EHT array. We apply this method to obtain a quantitative measurement of the brightness asymmetry in the images of \sgra. We then explore the impact on the brightness asymmetry of the black hole inclination and the velocity profile of the accretion flow simultaneously, for a wide range of black hole spins, to place constraints on the properties of \sgra\ and its accretion flow. We use the covariant semi-analytic model developed in \citet{BlackHoleImagesAsTestPlasma} to enable a variety of velocity profiles in the inner flow as well as to explore a broad range of characteristics. 

We begin by discussing the different methods of measuring the brightness asymmetry of black hole images and introducing a definition of image asymmetry in the Fourier domain in Section \ref{sec:asymmetry}. In Section \ref{sec:data}, we apply these methods to data from \sgra\ and M87 to measure the degree of asymmetry. Utilizing the measurements of Section \ref{sec:data} and the semi-analytic accretion disk and ray-tracing model employed by \cite{BlackHoleImagesAsTestPlasma}, we explore and provide constraints on inclination and Keplerianity of the accretion flow of the Sgr A* in Section \ref{sec:semianalytic}.

\section{Methods for Measuring Black Hole Image Asymmetry} \label{sec:asymmetry}

The EHT measures complex visibilities between various baselines of the global array, which are equal to the Fourier components of the black-hole image at particular spatial frequencies. Because of the large distances between the different stations in the array, these visibilities provide only a sparse coverage of the Fourier (also known as the $u-v$) plane. A variety of methods is then used to reconstruct images from the observed visibilities~\citep{SgrPaperIII, M87PaperIV}. In this section, we establish how brightness asymmetry can be defined in both the image and Fourier domains and introduce, for the latter, a new measure of brightness asymmetry in Fourier space that can be used to enable quantitative comparisons to models. 

\subsection{Image Domain Asymmetry}
\label{subsec:IA}

One method of measuring the brightness asymmetry for the black hole images was introduced by \citet{Medeiros_2022}. Following this method, we set a Cartesian coordinate system along the image plane with the black hole spin axis pointing along the positive y-axis. We center this coordinate system in the center of the black-hole shadow, which is shifted from the origin by $x = 2\, \text{a}\,  \text{sin}i$ along the axis perpendicular to the spin axis due to the effects of frame dragging (see \citealt{Bozza2006, JohansenCenterShift}).

Since the expected brightness asymmetry of the image due to the Doppler effects occurs at directions perpendicular to the orientation of the black-hole spin, we take a horizontal cross section of the image intensity along the $y=0$ line and integrate the intensity from the center of the shadow out to a radius $r_{\rm out}$ where the contributions to the flux are negligible. We repeat this integral towards the opposite direction and define the degree of asymmetry as the ratio of these two integrals minus one, i.e.,
\begin{equation}
    \text{IA}=\frac{\int_{0}^{r_{out}} dx\; I(x-2 a \sin i,y=0)}{\int^{0}_{-r_{out}} dx\; I(x-2 a \sin i,y=0)}-1
\label{eq:IA}    
\end{equation}
(Note that we alter the definition in \citealt{Medeiros_2022} by subtracting 1 from the ratio so that zero corresponds to a symmetric image).

\subsection{Fourier Domain Asymmetry}
\label{subsec:IA}

There are a number of drawbacks when measuring the image asymmetry through equation~(\ref{eq:IA}) or other image-domain based methods. First, we do not always know {\em a priori\/} the direction of the spin axis of the black hole. While some sources, such as M87, have constraints on the inclination and direction of the black hole spin axis based on the orientation of their jets \citep{Walker_Jets}, such constraints do not exist for \sgra. Second, the image asymmetry measure defined in equation (\ref{eq:IA}) does not capture all of the information in the image and focuses on the axis of greatest asymmetry. Third, the observational data are taken in the Fourier domain and converted to images through image reconstruction algorithms. In the image of \sgra, there are bright spots the locations of which change based on the various assumptions and regularizers used in those algorithms, indicating that they are most likely artifacts of the reconstruction. (They typically coincide with Fourier space locations where there are more data points.) These bright spots can alter the asymmetry measure in the image domain. To avoid a measure of asymmetry that is affected by specific image reconstruction algorithms, we introduce one that can be defined directly from the interferometric data.

The 2-D Fourier transform of an image is given by 
\begin{equation}\label{eq: 2DFT}
    V(u,v)=\iint dx dy I(x,y) e^{-2\pi (ux+vy)/\lambda},
\end{equation}
where $I(x,y)$ is the intensity along the image plane, $\lambda$ is the wavelength of the observation, and $V(u,v)$ is the complex visibility. It is these visibilities that are measured directly by the EHT.

\begin{figure*}[!ht!]
\includegraphics[width=0.95\textwidth]{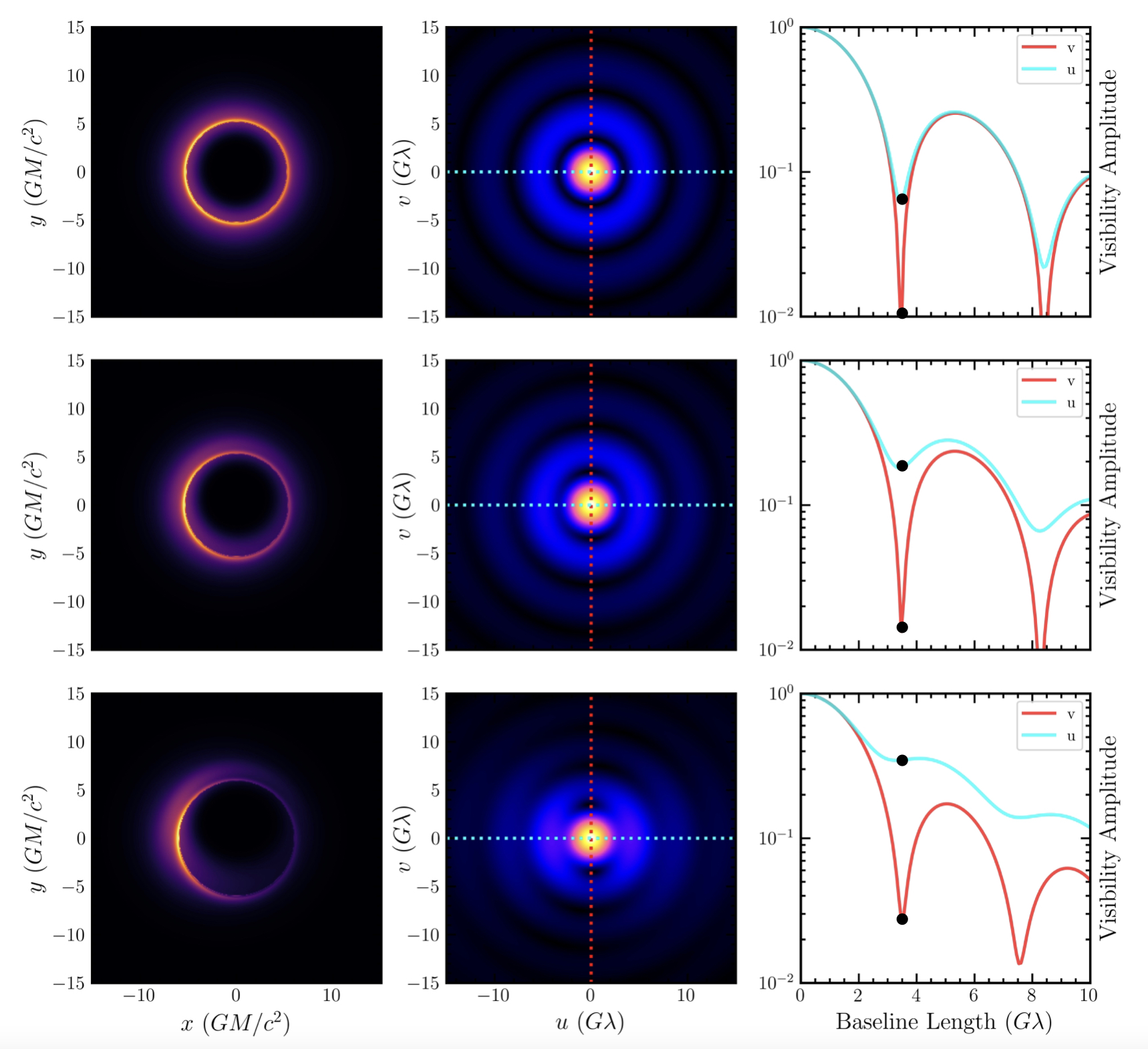}
\caption{\footnotesize (Left) Images generated from the semi-analytic model for black holes of spin $a=0$ and inclinations of 5$^{\circ}$, 15$^{\circ}$, and 30$^{\circ}$ from top to bottom. As the inclination increases, the image asymmetry also increases with inclination as expected. (Center) The visibility amplitude of each black hole images, with the orientations of the black hole spin axis and of the orthogonal axis plotted in red and cyan, respectively. (Right) Cross sections of the visibility amplitude along the two axes vs.\ baseline length. As the inclination increases, the amplitude of the first minimum along the $u-$axis increases drastically. The black hole images have been re-scaled so that the first deep minimum of each of the images are at the same baseline length. The black dots represent the relevant points along the two baselines for quantifying the Fourier asymmetry.}
\label{fig:uaxischangingasymmetry}
\end{figure*}

Figure~\ref{fig:uaxischangingasymmetry} illustrates a number of model images and their Fourier transforms, for black holes observed at different inclination angles and, hence, showing different degrees of Doppler asymmetry.  For a black hole with a spin axis along the positive $y$-axis (with the angular momentum of the accretion flow assumed to be in the same direction), the peak brightness of the image occurs on the left side of the shadow due to Doppler beaming, as can be seen in the left column of Figure~\ref{fig:uaxischangingasymmetry}. In the 2-D Fourier transform of completely symmetric, ring-like images (top panels), deep minima in the visibility amplitudes occur along both the u- and v-axis (or any arbitrarily defined axis through the transformed images). However, when Doppler beaming due to the azimuthal velocity of the accretion flow introduces an asymmetry along the $x$-axis in the image domain, the structure of the Fourier maps changes: deep minima still occur along the $v$-axis, i.e., the axis aligned with the black hole spin (see, e.g., the third column of Figure~\ref{fig:uaxischangingasymmetry}), but the minima along the $u$-axis, i.e., orthogonal to the black hole spin, become shallower. For example, the $30^{\circ}$ inclination model in the bottom panels of Figure~\ref{fig:uaxischangingasymmetry} has minima along the u- and v-axis that are significantly different in depth than (albeit at the same baselines as) those of the 5$^{\circ}$ model in the top panels.

We show in Figure~\ref{fig:uaxischangingasymmetry} these typical patterns for images and their Fourier transforms with increasing inclination (which is a proxy for increasing brightness asymmetry). We expect the smallest degree of asymmetry at the lowest observer inclination, where both the image and the Fourier transform appear highly symmetric, as in the top row. When the inclination increases (middle row), the first deep minimum along the axis aligned with the black hole remains in the figures of the visibility amplitude, but the values at this minimum along the axis orthogonal to the black hole spin axis are much shallower. These properties will be useful for defining a Fourier-domain asymmetry measurement. When the asymmetry is even higher (bottom row), the minima in the visibility amplitude in the orthogonal direction may not even be discernible.

We will use the first minimum in the visibility amplitude as our basis for the definition of asymmetry because it carries information about the primary features of the large-scale, ring-like structure of the black hole image. This aspect allows us to probe the global velocity profiles of the accretion flow with brightness asymmetry because Doppler boosting gives rise to  large-scale effects on the asymmetry of the image and does not impact significantly the smaller scale structures captured by longer-baseline data. Furthermore, both M87 and \sgra\ observations contain significant number of data points in this region of the $u-v$ plane, each with two different baselines approximately orthogonal to each other. This allows us to infer the degree of brightness asymmetry from the data, as the effects on the Fourier transform are most prominent at orientations that are 90$^\circ$ away from the baseline of the deep minimum. 

We introduce our measure of brightness asymmetry in the Fourier domain using simulated images. We first find the baseline length $b_{\rm min}$ of the first deep minimum of the Fourier transform of each image. In principle, we need to scan all orientations to find the direction of the deepest minimum, but, in practice, this is always aligned with the known black hole spin axis. We then find the visibility amplitude at this same baseline length but along a baseline that is oriented 90$^\circ$ away from the baseline corresponding to the deepest minimum. This orientation corresponds to the most asymmetric axis of the Fourier transforms. We then define the value of the Fourier-domain asymmetry as the ratio of the visibility amplitude along the asymmetric axis at the baseline corresponding to the deepest first minimum divided by the visibility amplitude at zero baseline length: 
\begin{equation}
    \text{FA}=\frac{|V(u=b_{\rm min},v=0)|}{|V(0,0)|}. 
\label{eq:FA}
\end{equation}
Here, $b_{\rm min}$ is the baseline length where the first minimum along the v-axis (or symmetric axis for an arbitrary image) occurs and $V(0,0)$ is the visibility amplitude at zero baseline length. 

In Appendix \ref{sec: Appendix A}, we compare this definition with a parameter that measures the asymmetry of crescents in the geometric model for black hole images devised by \citet{Crescent} and show that the two track each other monotonically. This further motivates and validates the use of Fourier asymmetry defined in equation~(\ref{eq:FA}). In the next section, we will use this method to quantify the degree of asymmetry in the black-hole images of \sgra\ and M87.

\section{Quantifying the Asymmetry of EHT Targets}
\label{sec:data}

Having established our definition of asymmetry of the black hole images in the Fourier domain, we now apply this measure directly to the 2017 visibility amplitude data of Sgr~A$^*$ and M87. In the case of M87, we also compare the measurements of both image-domain asymmetry (as reported in~\citealt{Medeiros_2022}) and Fourier-domain asymmetry to check for consistency between these two approaches. 

\begin{figure*}[!ht!]
\includegraphics[width=0.95\textwidth]{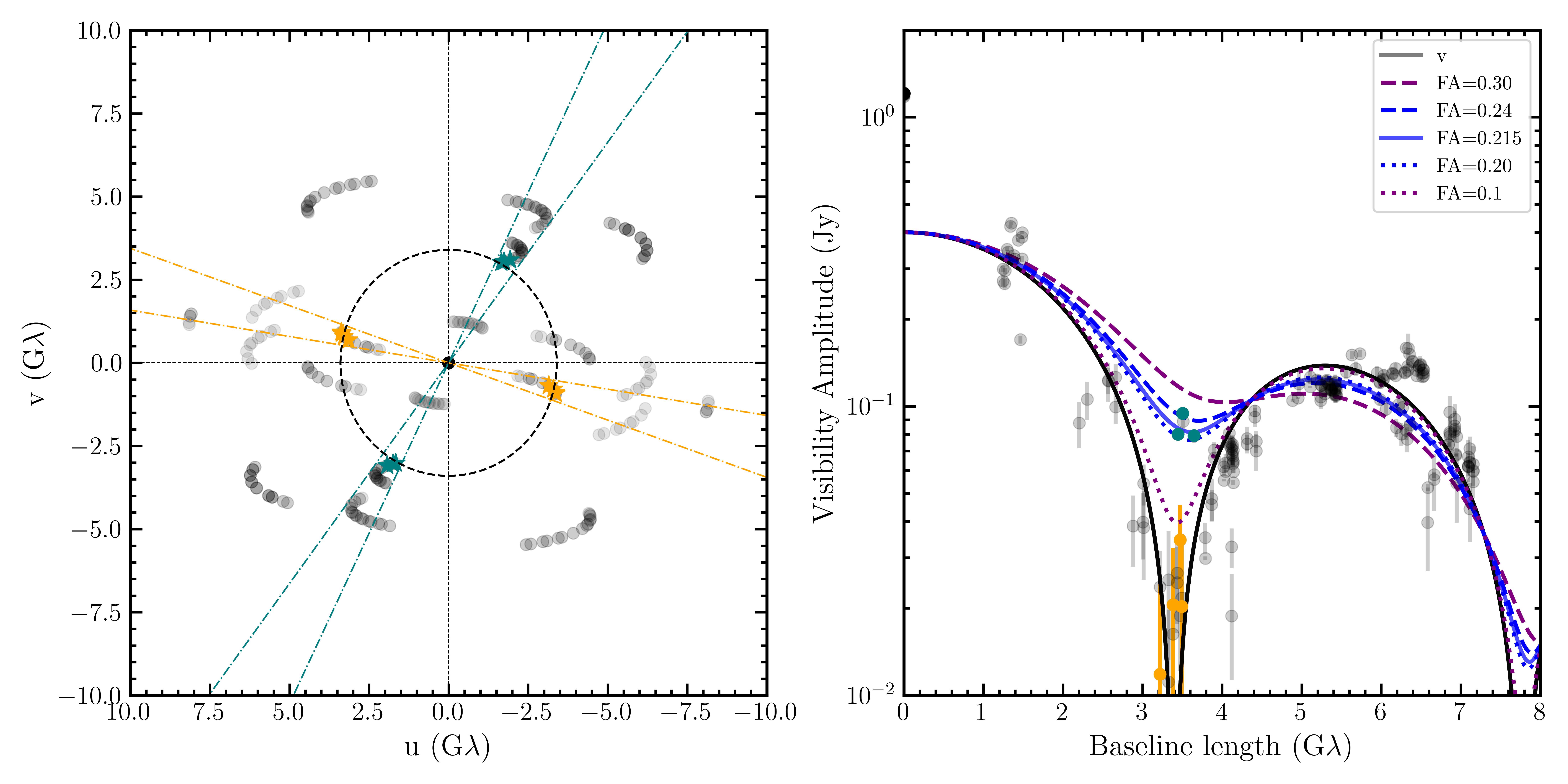}
\caption{\footnotesize (Left) The $u-v$ coverage for the M87 black hole using the 2017 April 11 EHT observations, where the data points of the two sets of baselines near the deep minimum used for Fourier asymmetry measure are shown in teal and orange and the remaining data points are shown in black. Dashed lines indicate the boundaries of the points used to measure the angle between the orientations of the two sets of baselines. (Right) The observed visibility amplitude versus baseline length for the same dataset, with the points used for the Fourier asymmetry shown in teal and orange. Plotted over the data are crescent models of constant thickness with varying degrees of symmetry, quantified by their Fourier asymmetry. Blue lines show the most likely crescent model and the range of acceptable models that agree with the data and are labeled by the corresponding values of Fourier asymmetry. The dotted purple lines show models with asymmetries that disagree with the measurements.}
\label{fig:M87}
\end{figure*}

We first analyze the observations of the M87 black hole taken on April 11, 2017 (see Figure \ref{fig:M87}). We follow the two requirements we outlined for the pairs of telescopes used for the asymmetry measurement. First, the baseline length between the two telescopes should be comparable to that of the first deep minimum. For M87, the baseline length of the first minimum is around $3.4\, G\lambda$ \citep{M87PaperI}. Second, measuring the Fourier asymmetry requires the orientation of the baseline corresponding to the deepest minimum in the visibility amplitude and of the second baseline to be nearly orthogonal to each other, for reasons discussed in the previous section. The telescope pairs that best fit these conditions for M87 are JCMT-LMT/SMA-LMT and APEX-LMT/ALMA-LMT, shown in orange and teal, respectively, in Figure~\ref{fig:M87}. While the angle between the orientations of the telescope pairs is $\sim 72.5^{\circ}$, which is less than the desired condition of orthogonality, this has a minor impact on the Fourier asymmetry measurement, as we show in Appendix~B.

We show in Figure~\ref{fig:M87} the visibility amplitude data and $u-v$ plane coverage of M87 \citep{M87PaperI}, and emphasize in colored data points the baselines we use for the symmetric axis containing the deep minimum (orange) and for the asymmetric axis (teal). We identify the range of points around the first minimum and use them to set lower and upper values of the Fourier asymmetry, calculated using Equation~(\ref{eq:FA}). This gives a lower value for the Fourier asymmetry of FA$=0.20$ and an upper value of FA$=0.24$. While there is some uncertainty in the zero-baseline flux for M87 because of the diffuse emission from the jets, we use a fiducial value of 0.4~Jy, which is consistent with values reported in \citet{M87PaperIV}. 

For illustration, we also show in the right panel of Figure~\ref{fig:M87} the Fourier transform of different crescent models~\citep{Crescent} that correspond to the measured values of Fourier asymmetry. (We choose the remaining parameters, such as the relative disk width $\psi$, the outer ring diameter, Gaussian blurring, etc., so that they are consistent with the crescent model fit results in \citealt{M87PaperVI}.) We use these analytic crescent models only to show as examples the profiles of visibility amplitude for models with the same degree of Fourier asymmetry as the data and do not fit them to the data. Additionally, we include, again for illustration purposes, crescent models with parameters corresponding to higher and lower asymmetry (e.g., FA$=0.1$ and FA$=0.3$), which are inconsistent with the M87 data. In Section~\ref{sec:semianalytic}, we will compare these measured values to the Fourier and image asymmetry calculated from the semi-analytic model of~\citet{BlackHoleImagesAsTestPlasma} to derive implications on the velocity of the plasma and the inclination of the accretion disk with respect to the observer. 

\begin{figure*}[ht]
\includegraphics[width=0.95\textwidth]{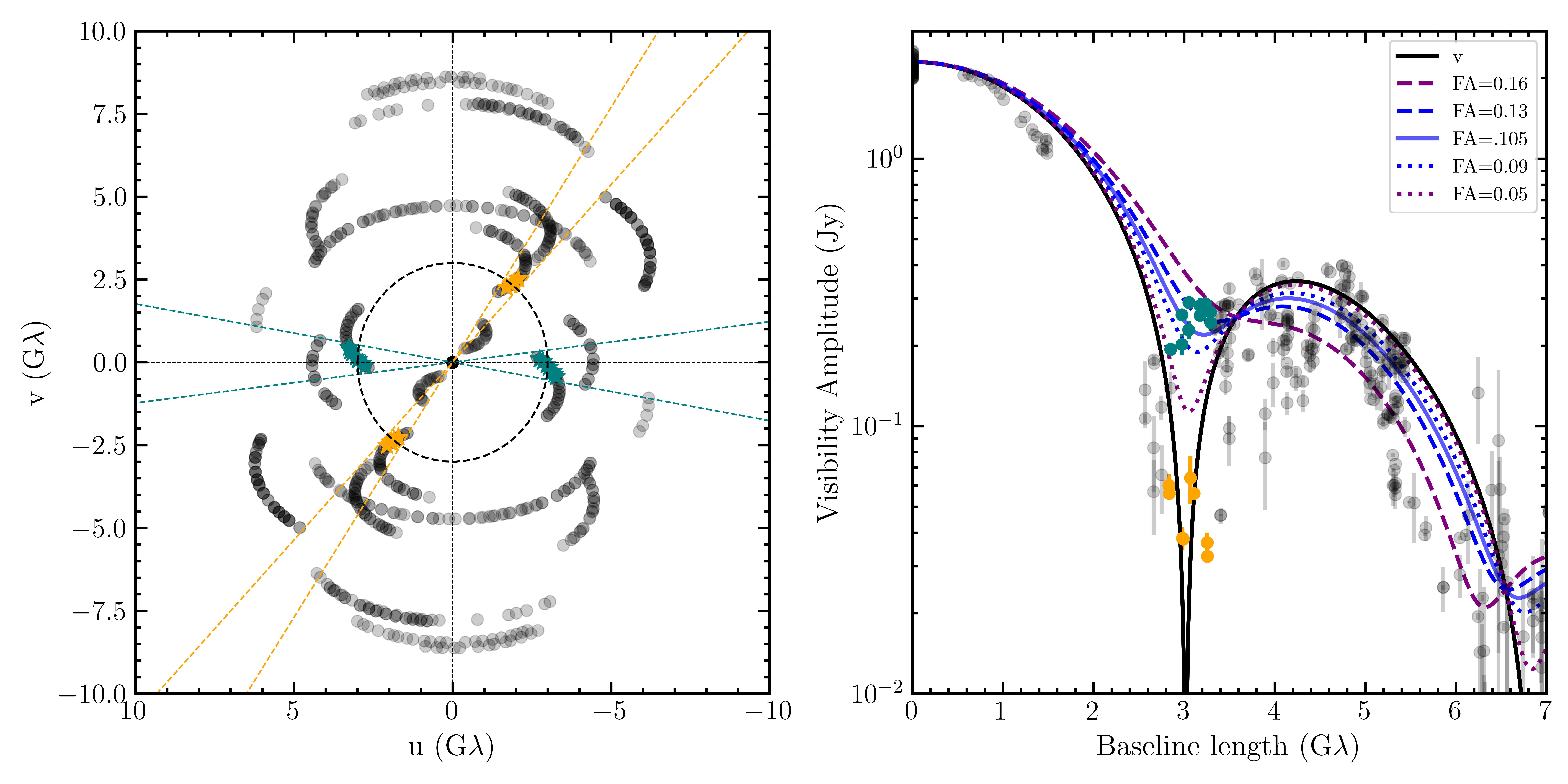}
\caption{\footnotesize Same as Figure~\ref{fig:M87} but for the 2017 April 7 EHT data of \sgra.}
\label{fig:Sgr}
\end{figure*}

We follow a similar procedure for analyzing the 2017 April 7 observations of Sgr~A$^*$. The two sets of baselines that best fit the requirements for quantifying the Fourier asymmetry for Sgr~A$^*$ are those of Chile$-$LMT and SMT$-$Hawaii, which have baseline lengths near the $3.0\,$G$\lambda$ minimum of the visibility amplitude \citep{SgrAPaperI}, as we show in Figure~\ref{fig:Sgr}. The angle between the orientations of the baseline pairs is $\sim 52.5^{\circ}$, which is again smaller than the desired orthogonality used for the definition of Fourier asymmetry in equation (\ref{eq:FA}), but this has a minor impact on the asymmetry measure (see Appendix \ref{sec: Appendix B}). 

We use the set of data points from these baselines to obtain a range of Fourier asymmetry values, following Equation~\ref{eq:FA}. Specifically, we take the visibility amplitude of the lowest and highest points marked in the right panel of Figure~\ref{fig:Sgr} and divide it by the zero-baseline flux, for which we use $2.3$~Jy. We obtain the range 0.09 to 0.13 for Fourier asymmetry. Comparing the plots of visibility amplitudes versus baseline length for the two EHT targets, we see that the asymmetry measured for Sgr~A$^*$ is significantly lower than that of M87. In the following section, we explore the implications of this surprisingly low asymmetry in the image of Sgr~A$^*$ for the accretion flow and the viewing geometry.

\section{Why is the image of Sgr~A$^*$ so symmetric?} 
\label{sec:semianalytic}

\subsection{Accretion Flow Model}
To explore potential reasons for the low degree of asymmetry in the EHT images of Sgr~A$^*$ and what that may imply about the properties of the accretion flow,  we use the covariant semi-analytic, thick-disk model for black hole accretion that was developed in \citet{BlackHoleImagesAsTestPlasma} and \citet{BlackHoleImagesAsTestSpacetime}. This model provides a highly flexible framework for exploring a broad range of plasma conditions and black hole properties, enabling a variety of velocity and density profiles in the accretion flow in general spacetimes while self-consistently determining the flow properties using basic conservation laws. We summarize here the main elements of this model and refer the reader to \citet{BlackHoleImagesAsTestPlasma} for a complete description. 

In this model, we solve for the fluid density $\rho$, the four-velocity $u^\mu$, the magnetic field strength $B$, and ion temperature $T_i$ using the conservation equations for the stress-energy tensor of the plasma, $T^{\mu\nu}$, and the energy-momentum of the flow, $\rho u^\mu$. We assume axisymmetry such that there is no dependence of any quantity on the azimuthal angle $\phi$ and calculate vertically averaged quantities for the flow. We consider a radiatively-inefficient flow with a stress-energy tensor that primarily includes viscous and compressional stresses. We obtain the normalization of the electron (ion) density from the mass accretion $\Dot{M}$, as a function of the radial fluid velocity $u_r$ and the scale height of the disk $h/r$. Given the calculated properties of the flow, we obtain the emissivity of the plasma assuming only optically thin, synchrotron emission from a thermal, isotropic distribution of relativistic electrons. We use the analytic fitting formula from \cite{Synchrotron} for synchrotron emissivity.

The plasma exhibits both azimuthal and radial motion. Plasma elements orbit the black hole with velocities comparable to the geodesic angular velocities of test particle as well as drift inward as a result of the outward transport of angular momentum. The profile of the radial velocity, which depends on the various stresses and rate of angular momentum transport as well as on the role of pressure support, is specified as a general power law form outside the radius of the innermost stable circular orbit (ISCO) with two free parameters that describe the magnitude and index of the inward velocity of the plasma flow. On the equatorial plane, it is given by
\begin{equation}
    u^r_{eq}(r)=-\eta \left(\frac{r}{r_{\rm ISCO}}\right)^{-n_r}\;.
\end{equation}
We use $\eta=0.1$ and $n_r=1.5$, unless otherwise stated, which have been shown to describe the velocity profiles obtained by recent GRMHD simulations~\citep{BlackHoleImagesAsTestPlasma}. 

In the model presented by \citet{BlackHoleImagesAsTestPlasma}, the azimuthal plasma velocities are equal to those of Keplerian orbits of test particles on the equatorial plane down to the ISCO, at which point the orbits become unstable and the plasma plunges toward the horizon following geodesics. In this paper, because the azimuthal velocity has a large impact on the images and, in particular, Doppler beaming is expected to be the primary cause of asymmetry in the images of black holes, we relax this assumption and investigate the potential for the angular velocity of the plasma flow to be sub-Keplerian. 

The azimuthal plasma velocities may become sub-Keplerian because of radial pressure gradients that can provide centrifugal support to the flow. The Advection Dominated Accretion Flow solution of \citet{NarayanYiADAF} finds, for this reason, angular velocities that are only a fraction of the local Keplerian velocities. Alternatively, magnetic field stresses in magnetically arrested disks (MAD) could also slow the angular rotation by facilitating efficient transport of angular momentum. Numerical simulations of MAD disks have indeed shown equatorial, azimuthal velocities that are also only a fraction of the Keplerian velocity (e.g., \citealt{Medeiros_2022}). 

To allow for sub-Keplerian velocity profiles, we introduce the phenomenological parameter $\epsilon$ such that the angular velocity of a test particle observed by an observer at infinity becomes \citep{Ryan1995}
\begin{equation}
    \Omega(r)= \epsilon  \frac{-g_{t\phi,r} + \sqrt{(g_{t\phi,r})^2 - g_{tt,r}g_{\phi\phi,r}}}{g_{\phi\phi,r}}
\end{equation}
and the equatorial azimuthal velocity, as seen from an observer at infinity, outside of the ISCO can be calculated as 
\begin{equation}
    u_{eq}^{\phi}(r)=\frac{\Omega}{\sqrt{-g_{tt} - (2g_{t\phi}+g_{\phi\phi}\Omega)\Omega}}. 
\end{equation}
The lower limit of the parameter $\epsilon$ is set such that, for each value of the black-hole spin, the $\phi$ component of the observed angular velocity does not lead to unphysical velocity four-vectors due to frame dragging.

With the $\theta$ component of the four-velocity set to zero by assumption, the $t-$component can also be found by the condition that the norm of the four-velocity be equal to -1. At the ISCO, the energy and angular momentum of the particles are calculated using the prescription for $r>r_{\rm ISCO}$, which are then used to calculate the $\phi-$ and $t-$ components of velocity of the plasma inside the ISCO under the assumption that the plasma loses centrifugal support and plunges towards the event horizon. 

Off of the equatorial plane, the results of semi-analytic models of geometrically thick flows (such as \citealt{NarayanYi}) and those of GRMHD simulations (e.g., \citealt{Sadowski}) show that azimuthal velocities are constant along spherical radii and the radial velocities are constant along cylindrical radii. We use these conditions to specify the velocities off of the equatorial plane. 

Having calculated the accretion disk properties such as the density, magnetic field, and temperature as well as the velocities of the flow, we then perform relativistic ray-tracing to integrate the radiative transfer equation along geodesics from the observer to the accretion flow using the algorithm of \citet{PsaltisRay}.  

\subsection{Application to EHT}

Using the model described in the previous section, we now explore image properties for a broad range of black hole spins, observer inclinations, and velocity profile parameters, focusing specifically on large-scale asymmetries in the images. We first obtain a relationship between image asymmetry and Fourier asymmetry, both of which are easily calculated from the semi-analytic models, in order to facilitate a direct comparison with observations. We then use this relationship to investigate the impact of sub-Keplerianity in the flow of \sgra\ by comparing the measurements of Fourier asymmetry from the observational data and the image asymmetry calculated with the semi-analytic models. 

\begin{figure}[t]
\includegraphics[width=1.08\columnwidth]{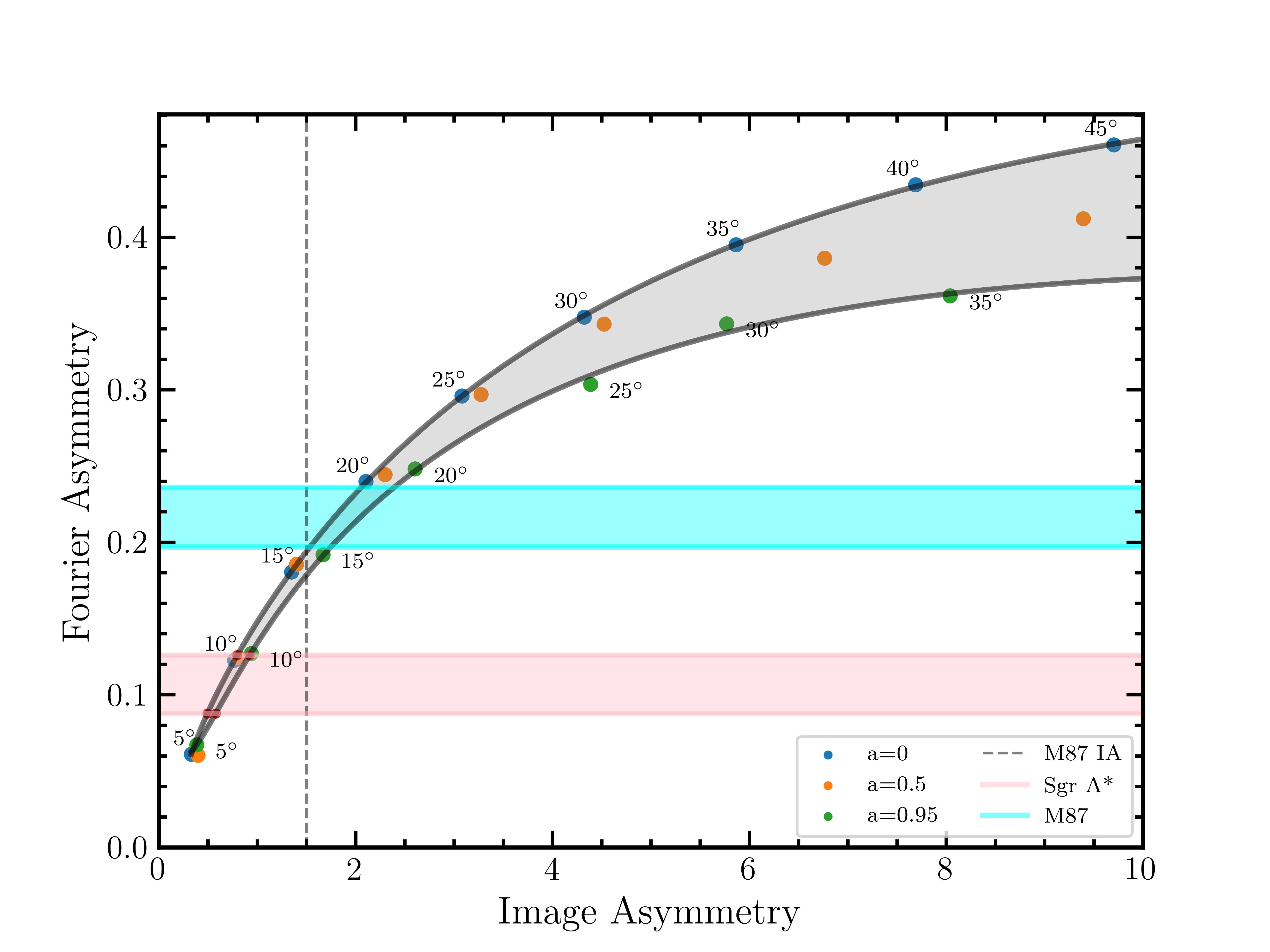}
\caption{\footnotesize The correspondence between the large-scale asymmetry measured in the image domain (Eq.~\ref{eq:IA}) and in the Fourier domain (Eq.~\ref{eq:FA}). The sets of points show the asymmetry calculated for semi-analytic models with Keplerian velocities ($\epsilon=1$) and different black hole spins and inclinations. There is a small increase in image asymmetry with increasing spin that is dwarfed by the effect of increasing inclination. The dotted black line shows the asymmetry measured in the image domain for M87 \citep{Medeiros_2022}, which matches the $\sim 17^\circ$ inclination inferred from its jets~\citep{Walker_Jets}. Cyan lines show lower and upper bounds for the Fourier asymmetry measurements from the visibility amplitude data of M87, while the pink lines show the same for Sgr~A$^*$.}
\label{fig:FAvsIA}
\end{figure}

Figure~\ref{fig:FAvsIA} shows the results from a series of semi-analytic models for black hole spins of $a=$0, 0.5, and 0.95 and observer inclinations ranging from 5$^{\circ}$ to 45$^{\circ}$, for accretion flow velocities that are purely Keplerian. For each of these models, we generate the image and calculate the value for image asymmetry according to Equation~(\ref{eq:IA}). We then take a 2-dimensional Fourier transform of each image and calculate the Fourier asymmetry using Equation~(\ref{eq:FA}). The tight correlation between the Fourier asymmetry and image asymmetry for these models for the three black hole spins illustrate the weak dependence of these two asymmetries introduced by the unknown spins. 

The relationship shown in Figure~\ref{fig:FAvsIA} allows us to assign an effective inclination to the EHT images assuming that the $\phi$-component of the velocity in the accretion flow is Keplerian. For the M87 black hole, we find an inclination between 15$^{\circ}$ and 20$^{\circ}$, which is consistent with that found from relativistic jets \citep{Walker_Jets}. This is also consistent with the image brightness asymmetry found by \citet{Medeiros_2022} and shown as a dashed vertical line in Figure~\ref{fig:FAvsIA}.

When we apply the same technique to Sgr~A$^*$, we find that the observer's inclination is confined to 6$^{\circ}-10^{\circ}$. This extremely small inclination raises some concerns from both physical and probabilistic perspectives. The chance that \sgra\ is pointing towards the Earth within $6-10^{\circ}$ from our line-of-sight compared to the entire solid angle of the sky is quite small (0.75\% for the $10^{\circ}$ case). This concern is exacerbated by the fact that the inclination of the other observable black hole M87 is also low ($17^{\circ}$). Given that our line-of-sight to Sgr~A$^*$ is within the disk of the Galaxy and is perpendicular to the angular momentum axis of the galaxy, this configuration for Sgr~A$^*$ would imply that the spin axis of the black hole is nearly orthogonal to the angular momentum axis of the Galaxy. Given the unlikeliness of this configuration, we explore an alternative effect that may cause this lack of image asymmetry, namely that the accretion flow for Sgr~A$^*$ could be significantly sub-Keplerian, which could have implications for the physics of the accretion flow. 

We utilize the sub-Keplerian parameter, $\epsilon$, to explore the effects of angular velocity on the image characteristics, most notably the asymmetry of the images. As Doppler beaming is the major cause of the large-scale image asymmetry of the black holes and because both inclination and sub-Keplerianity affect the degree of Doppler boosting, we explore these parameters in tandem. Doppler boosting depends only on the component of velocity along the line of sight to the observer. This is the reason why the smaller the inclination of the source, the lower the magnitude of the component of velocity toward the observer and, hence, the smaller the brightness asymmetry between the two sides of image. Introducing a parameter that decreases the $\phi$-component of the velocity of the flow also reduces the brightness asymmetry of the image as the relativistic beaming scales with the velocity of the emitting material. 

\begin{figure}[ht]
\includegraphics[width=0.9\columnwidth]{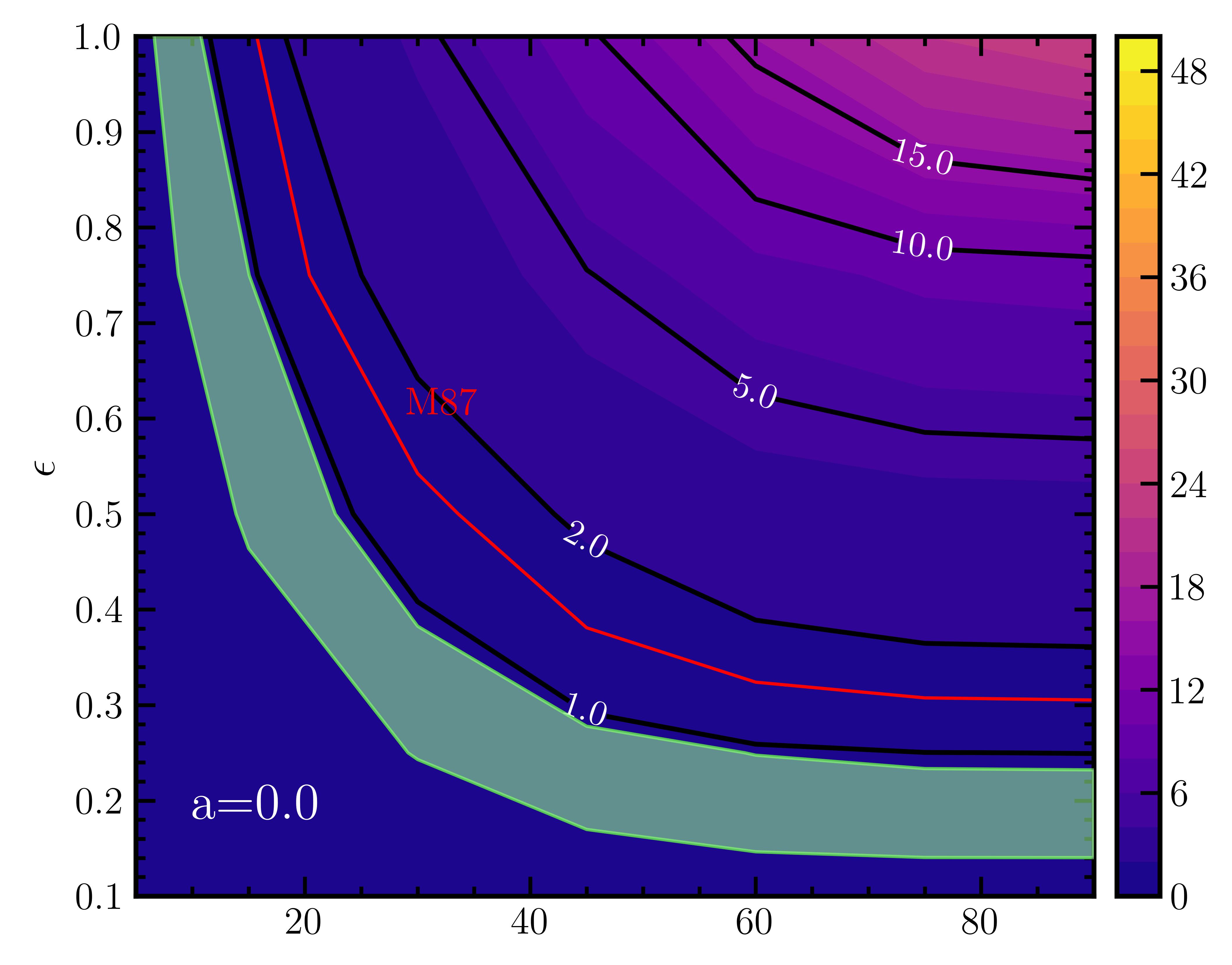}
\includegraphics[width=0.9\columnwidth]{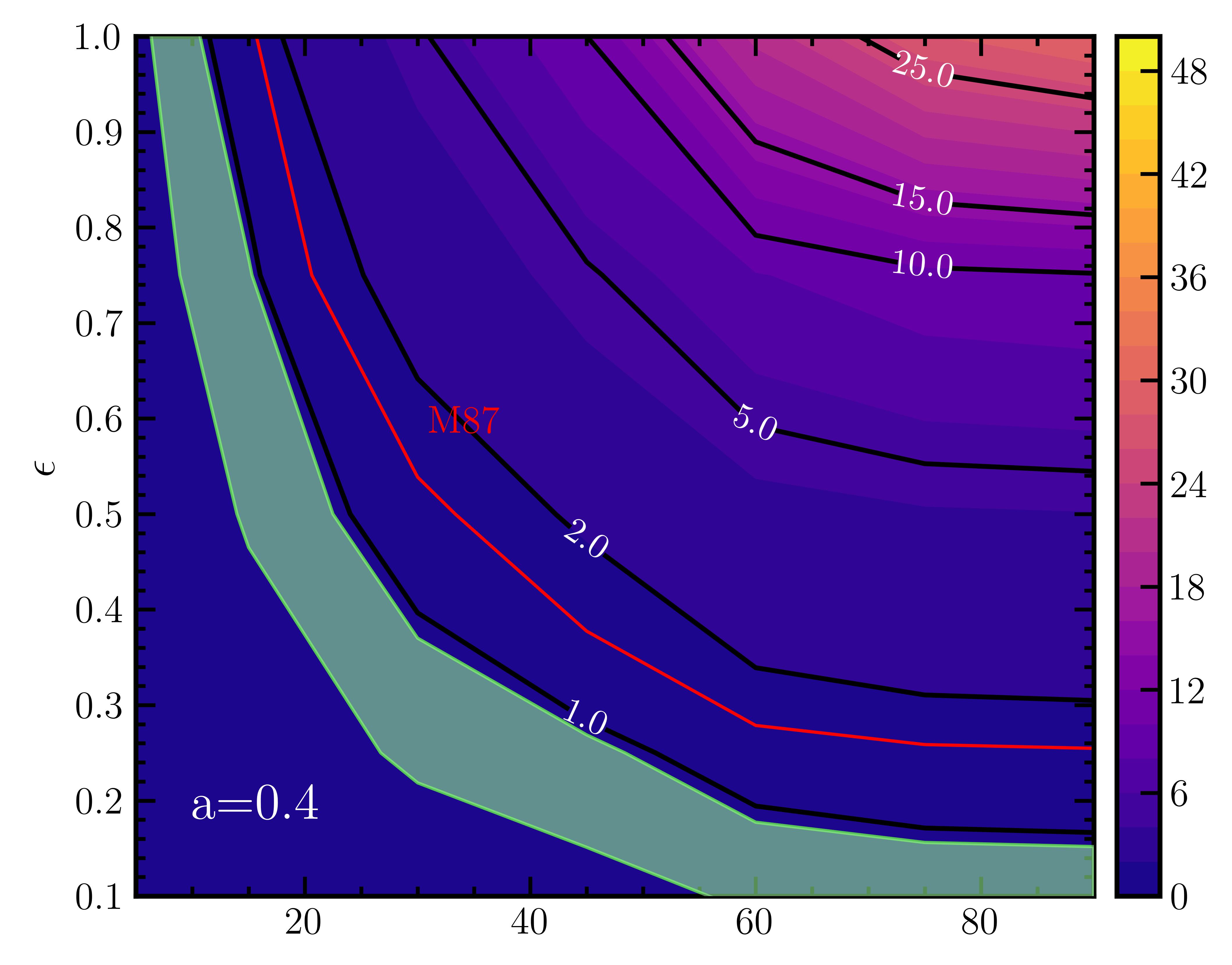}
\includegraphics[width=0.9\columnwidth]{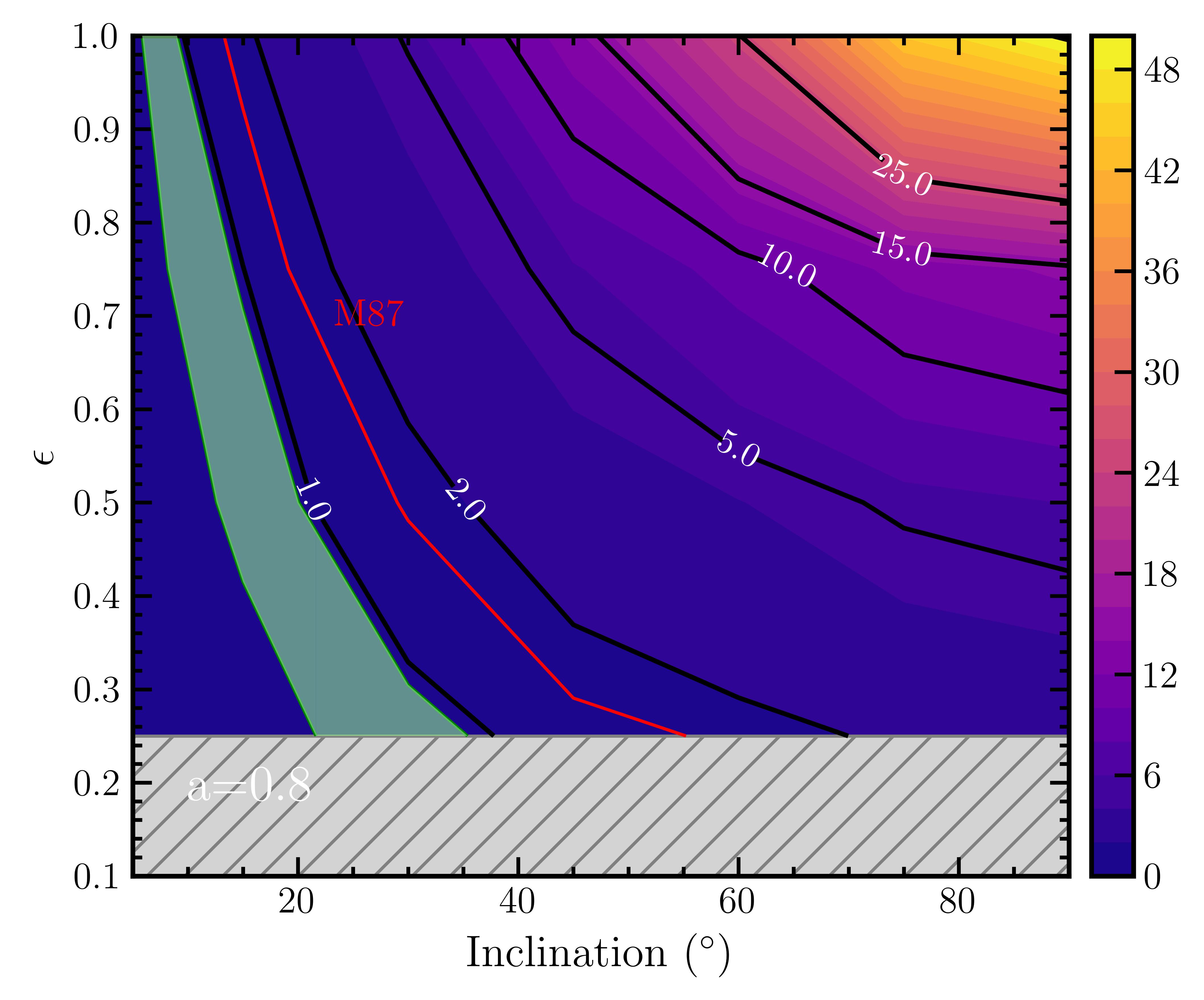}
\caption{\footnotesize Contours of image asymmetry, varying the sub-Keplerian parameter $\epsilon$ and observer inclination $i$, calculated from semi-analytic models for black holes with (top) spin $a=0$, (middle) $a=0.4$, and (bottom) $a=0.8$. The red line shows the measured image asymmetry for the M87 black hole, while the green lines and the shaded region represent the contours of image asymmetry for Sgr~A$^*$. The grey, hatched region in the bottom panel indicates the region where such a small angular velocity at infinity is impossible due to the effects of frame dragging.}
\label{fig:a0}
\end{figure}

The three panels of Figures~\ref{fig:a0} show contours of constant image asymmetry calculated from the semi-analytic model using different combinations of inclinations and sub-Keplerianity parameters $\epsilon$, for three values of black hole spin. If we ignore our prior knowledge of the inclination of the M87 black hole, we can evaluate the consistency of Keplerian angular velocities with the brightness asymmetry measured in that source. Additionally, in these figures, we use the relationship we found in Figure~\ref{fig:FAvsIA} to translate the measurements of Fourier asymmetry from Section~\ref{sec:data} into image asymmetry bounds and show them in the green band for Sgr~A$^*$. These contours  determine the combinations of inclination and sub-Keplerianity that are required to produce the observed asymmetry in \sgra. We note that, at high black-hole spins, there are regimes where very low values of angular velocity, corresponding to low values of $\epsilon$, lead to physical inconsistencies within the model, indicated by the grey hatched region of the bottom panel. This is because the angular velocity seen by an observer at infinity cannot be set arbitrarily low without conflicting with the effect of frame dragging. 

As expected, in all three panels of Figure~\ref{fig:a0}, the assumption of a Keplerian velocity profile for the accretion flow is consistent with the observed low value of image asymmetry only for very small observer inclinations. For the M87 black hole, the observed weak asymmetry indicates that Keplerian or near-Keplerian angular velocities are consistent with the known low inclination of this source. For \sgra, a value for the sub-Keplerian parameter that is significantly away from unity is required for inclinations that increase toward edge-on configurations. Furthermore, the value of the sub-Keplerian parameter that is required to accommodate a particular inclination becomes smaller as the black hole spin increases. For instance, for a black hole of spin $a=0$, the requirement for sub-Keplerianity associated with $i>45^\circ$ is $\epsilon \lesssim 0.3$, whereas this value falls below $\epsilon \lesssim 0.2$ for even a spin of $a=0.4$, as seen in the middle panel of Figure~\ref{fig:a0}.

We conclude from these results that a significant degree of deviation from Keplerian angular velocities is required to account for the degree of symmetry in the image of Sgr~A$^*$ in order to avoid a nearly pole-on viewing inclination. Those small values of the angular velocity could be indicative of, e.g., a large degree of pressure support or of magnetic stresses in the accretion flow. Even with that condition, a low spin for the black hole itself appears to be favored to make it possible for the inclination to be $\gtrsim 30^\circ$.  

\section{Discussion}  
In this paper, we introduced a new method for quantifying the asymmetry of black hole images based directly on interferometric data obtained with the EHT. This method is robust against potential artifacts that arise from the methodology or assumptions made to reconstruct images. The resulting measure of Fourier asymmetry utilizes the depth of the first deep minimum in the visibility amplitude of ring- and crescent-like images and its dependence on the degree of image asymmetry. 

We then applied this measurement to EHT data for \sgra\ and the black hole in M87 to quantify the degree of asymmetry in both sources. We find that both sources show low degrees of brightness asymmetry. The lack of significant brightness asymmetry implies low observer inclinations under the assumption of Keplerian or near-Keplerian velocity profiles. For M87, this is consistent with the 17$^{\circ}$ inclination inferred from observations of its jet. However, for Sgr A$^*$, we also find a very small inclination of $i\le 10^{\circ}$. While there are some potential pitfalls in interpreting posteriors of chance occurrence, we can calculate the likelihood that both of the EHT black hole targets are pointed towards us within these inclinations (i.e., $\lesssim 6-10^{\circ}$ for \sgra\ and $\lesssim 17^{\circ}$ for M87) to be equal to $(6-16)\times 10^{-5}$. This is uncomfortably low.

For \sgra, such a low inclination also implies a nearly orthogonal configuration between the spin axis of the supermassive black hole and the angular momentum of the galaxy as well as no alignment with the angular momentum of the nearby stellar population \citep{BartkoSgrAStellar}. Although stochastic events can impact the angular momentum and spin of the black hole, galaxy growth models involving stochastic events such as mergers still predict some correlation from coevolution of the central black holes and their surrounding galaxies \citep{BlackHoleCoevolutionReivew}.

There is no physical argument that requires angular velocities in accretion flows to follow a Keplerian profile. We showed that the relaxation of this assumption also relaxes the inclination constraints on the black hole. Making the velocities of the orbiting plasma to be sub-Keplerian requires forces in the accretion disk comparable to that of gravity that either provide significant centrifugal support, such as radial pressure gradients, or remove angular momentum, such as magnetic stresses. Albeit plausible, it is worth noting that the presence of such forces, which are highly stochastic, will likely exacerbate the tension between the predicted and observed variability of the 1.3~mm flux generated near the black hole horizon.

\appendix

\section{Correspondence between the Fourier Asymmetry and Geometric Crescent Model Asymmetry}
\label{sec: Appendix A}

We compare the Fourier asymmetry measure we introduced in this paper to the analytic crescent model of \citet{Crescent}, which has been used to characterize and fit black hole images. Our goal is to demonstrate the direct relationship between the Fourier asymmetry and the crescent model asymmetry quantified by the parameter $\tau$, as we discuss below.

\begin{figure}[t]
\includegraphics[width=0.95\columnwidth]{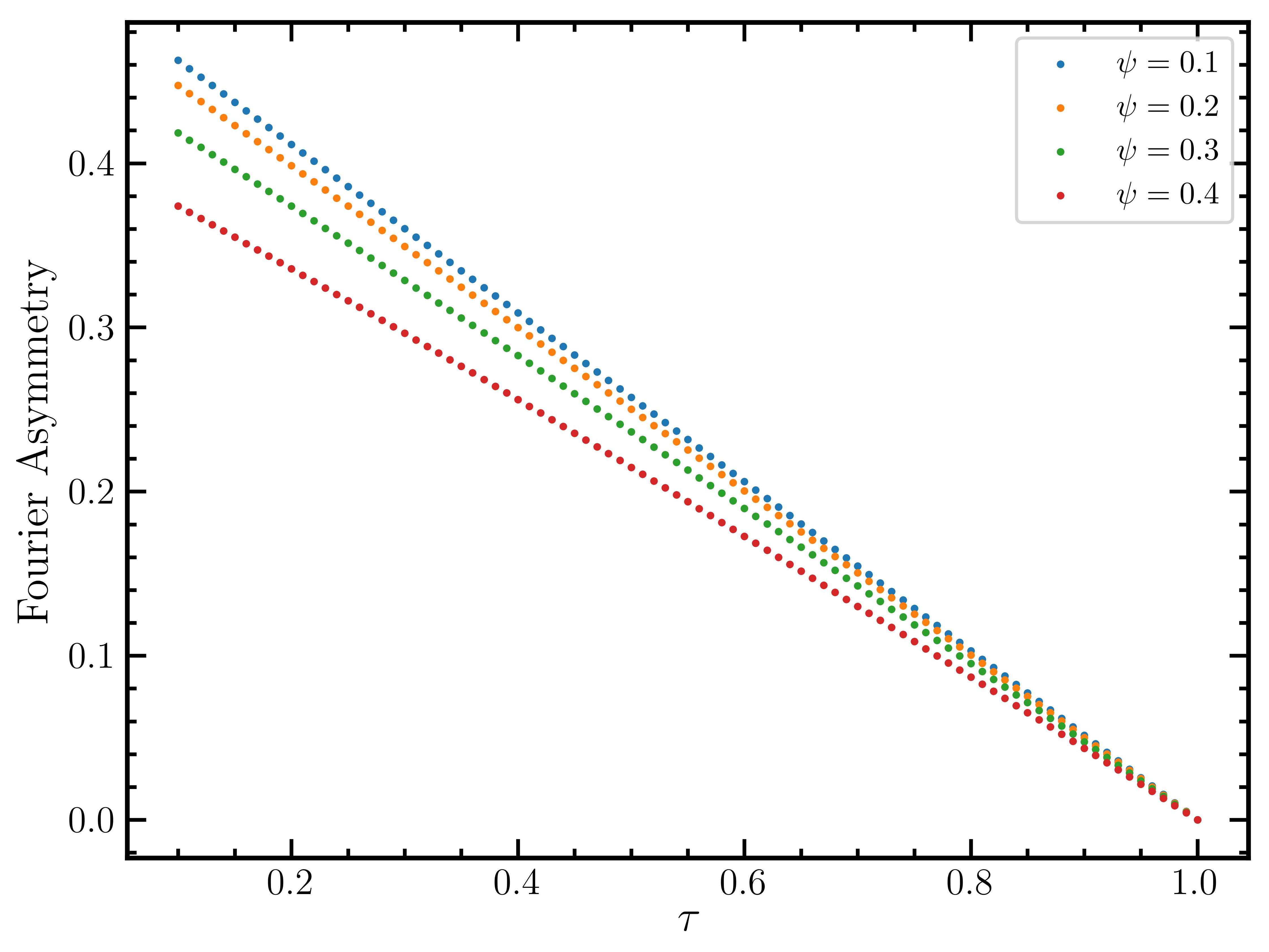}
\caption{Measurement of Fourier asymmetry for crescent models of different relative thickness plotted against the geometric measurement of asymmetry $\tau$. The value $\tau$=1 corresponds to the inner and outer disks of the crescent model are concentric. A smaller value of $\tau$ results in a more asymmetric image.}
\label{fig:tauvsfa}
\end{figure}

The crescent  model is a flexible geometric prescription that generates crescent images by subtracting an offset smaller disk (with radius $R_2$) from a larger disk (with radius $R_1$), both of constant intensity. We define the relative thickness of the crescent as 
\begin{equation}
\psi \equiv 1- \frac{R_1}{R_2}
\end{equation}
and the degree of asymmetry of the crescent as 
\begin{equation}
\tau\equiv 1- \frac{a}{R_1-R_2}\;.
\end{equation}
Here, $a_0$ is the x-offset of the center of the inner ring from the origin. In our implementation, we consider the center of the inner disk being offset along the x-direction, as the image and 2-D Fourier transform can always be rotated without loss of generality. 

We can construct the 2-D Fourier transform of a crescent by subtracting the Fourier transforms of the two disks from each other. Following the prescription by \citet{Crescent}, the complex visibility of the crescent is given by 
\begin{equation}\label{eq:crescentvisibility}
        V(u,v)= \frac{R_1}{2 \pi k (R_1^2-R_2^2)}[J_1(R_1 k) -   e^{-2\pi i a_0 u} \frac{R_2}{R_1} J_1(k R_2) ]
\end{equation}
where u and v are the spatial frequencies measured in the wavelength of observation, $k\equiv\sqrt{u^2 + v^2}$, and  $J_1$ is the first Bessel function of the first kind. 

\begin{figure}[t]
\includegraphics[width=0.99\columnwidth]{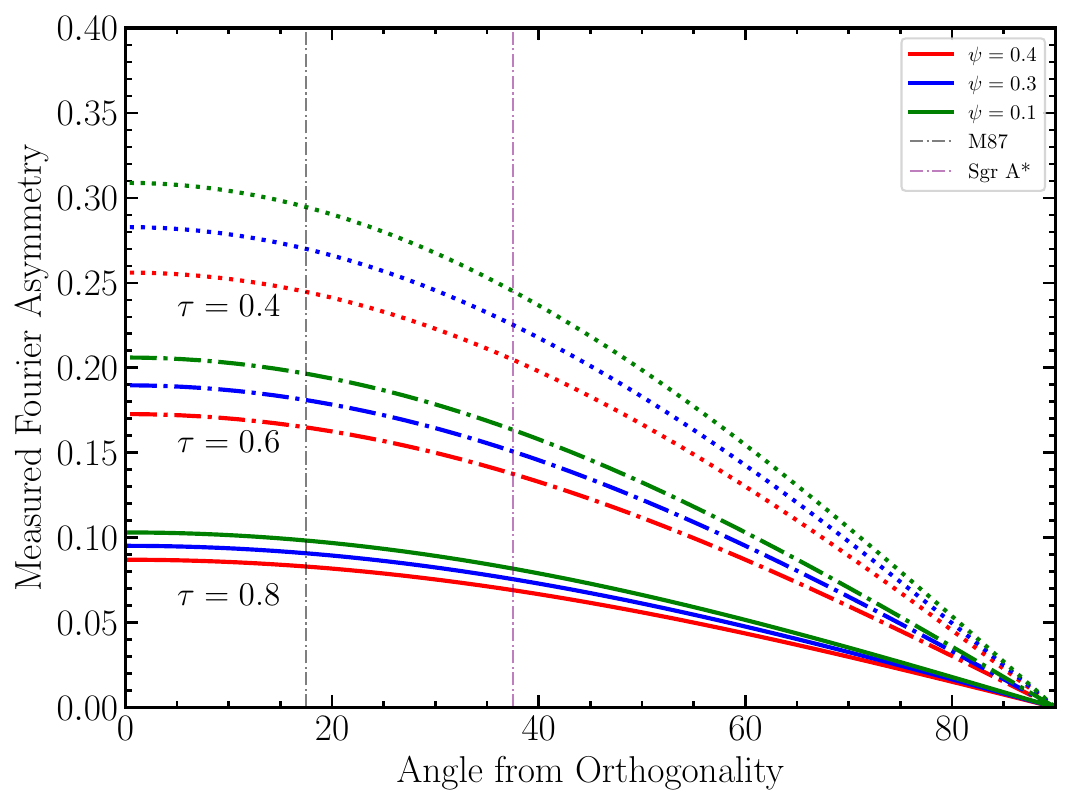}
\caption{Dependence of the measured Fourier asymmetry  on the angle between the axis in the Fourier map where the measurement is made and the orthogonal orientation to the black hole spin axis. The line colors and types represent different values of the model parameters. The vertical lines correspond to the orientations along which the asymmetry measurement was made for the M87 black hole and for \sgra\ (see Figs.~\ref{fig:M87} and \ref{fig:Sgr}).}
\label{fig:angle}
\end{figure}
 
Figure~\ref{fig:tauvsfa} shows the relationship between the Fourier asymmetry measurement of the crescent models (obtained by applying eq.~[\ref{eq:FA}]) against the geometric model asymmetry parameter $\tau$, for several values of $\psi$, the relative thickness of the crescent. We can see that, for higher values of $\tau$, indicating smaller asymmetry, the lower the  measured value of Fourier Asymmetry is. This is consistent with our expectation, as $\tau=1$ corresponds to zero offset of the center of the inner ring from the origin, resulting in a fully symmetric image. The figure also shows that there is a weak dependence on the relative thickness $\psi$ of the crescent for the same geometric asymmetry, with lower values of $\psi$ (thicker crescents) corresponding to slightly lower values of Fourier asymmetry. This result demonstrates a direct relationship between the Fourier asymmetry value defined in equation~(\ref{eq:FA}) and the geometric representation of asymmetry in the crescent models.

\section{Dependence of Fourier Asymmetry on the Angle between Baselines}
\label{sec: Appendix B}

The definition of the Fourier asymmetry in Equation~(\ref{eq:FA}) relies on using the visibility amplitude at a baseline that probes the first deep minimum and a second baseline that is oriented 90$^{\circ}$ away from it. However, even though there are baselines in the EHT data that probe the first minimum, the orientations of other available baselines are not exactly orthogonal to them. In this Appendix, we utilize the crescent model by \citet{Crescent}
to determine the effect of the relative orientation of the baselines on the measured Fourier asymmetry. 

Figure~\ref{fig:angle} shows the dependence of the measured value of Fourier asymmetry on the relative angle between the orientations of the two baselines measured for a set of crescent models with different values of asymmetry $\tau$ and for different crescent thicknesses. As expected, the maximum value of Fourier asymmetry is obtained when the two axes are orthogonal, represented by zero degrees on the x-axis. Vertical dashed lines show the relative orientations of the baselines that probe the first minimum for \sgra\ and M87. When we vary the angle of the baseline that corresponds to the shallow minimum away from the 90$^{\circ}$ orientation, the difference in the Fourier asymmetry does not change by more than 0.04, or 10\%, up to orientations that are $\sim 40^\circ$ away from orthogonality (for \sgra) even for the most asymmetric case, shown as $\tau=0.4$ in Figure~\ref{fig:angle}. The change in the Fourier asymmetry is even lower for models with lower crescent asymmetry. As a result, we can reliably obtain an accurate value for the Fourier asymmetry even when the baselines are not orthogonal to each other, as long as one of the baselines probe the location of the first deep minimum in the visibility amplitude and a close-to-orthogonal second baseline exists for the same baseline length. 

\bibliography{BH_asymmetry}{}
\bibliographystyle{aasjournal}

\end{document}